\documentclass[%
 reprint,
 amsmath,amssymb,
 aps,
 prl,
 longbibliography,
 lengthcheck,%
]{revtex4}

\usepackage{graphicx}
\usepackage{dcolumn}
\usepackage{bm}
\usepackage{hyperref}

\begin{document}

\preprint{APS/123-QED}

\title{Thickness-dependent magnetic properties of oxygen-deficient
EuO}

\author{M. Barbagallo$^1$}
\email{massimokriya@gmail.com}
\author{T. Stollenwerk$^2$}
\author{J. Kroha$^2$}
\author{N.-J. Steinke$^1$}
\author{N.D.M. Hine$^{1,3}$}
\author{J.F.K. Cooper$^1$}
\author{C.H.W. Barnes$^1$}
\email{chwb101@cam.ac.uk}
\author{A. Ionescu$^1$}
\author{P.M.D.S. Monteiro$^1$}
\author{J.-Y. Kim$^1$}
\author{K.R.A. Ziebeck$^1$}
\author{C.J. Kinane$^4$}
\author{R.M. Dalgliesh$^4$}
\author{T.R. Charlton$^4$}
\author{S. Langridge$^4$}
\affiliation{$^1$ Cavendish Laboratory, Physics Department,
University of Cambridge, Cambridge CB3 0HE, United Kingdom\\
$^2$ Physikalisches Institut and Bethe Center for Theoretical Physics,
Universit{\"{a}}t Bonn, D-53115 Bonn, Germany\\
$^3$ Thomas Young Centre, Department of Materials and Department of Physics,
Imperial College London, Exhibition Road SW7 2AZ, United Kingdom\\
$^4$ ISIS, Harwell Science and Innovation Campus, STFC, Oxon OX11 0QX,
United Kingdom}

\date{\today}

\begin{abstract}
We have studied how the magnetic properties of oxygen-deficient EuO sputtered
thin films vary as a function of thickness. The magnetic moment, measured by
polarized neutron reflectometry, and the Curie temperature are found to
decrease with reducing thickness. Our results indicate that the reduced number of nearest neighbors, band bending and the partial depopulation of the electronic states that carry the spins associated with the 4\textit{f} orbitals of Eu are all contributing factors in the surface-induced change of the magnetic properties of EuO$_{1\mbox{-}x}$.
\begin{description}
\item[PACS numbers:]
75.70.-i, 75.50.Pp, 75.47.Lx, 71.15.Mb
\end{description}

\end{abstract}

\maketitle

Electron-doped EuO is a semiconductor which undergoes
a simultaneous ferromagnetic and insulating-conducting phase transition, across which the resistivity drops by 8 to 13 orders of magnitude \cite{Oliver72,Penney72} and the conduction electrons become nearly 100 \% spin polarized \cite{Steeneken02,Ott06}, making EuO a strong candidate for efficient spin filtering \cite{Schmehl,Barbagallo}. Electron doping increases the Curie temperature of EuO thin films to above 200 K \cite{Miyazaki} from 70 K for undoped EuO, and
also increases the magnetic moment up to 7.13 $\mu_{\textsc{B}}$ from the
intrinsic value of 7 $\mu_{\textsc{B}}$ \cite{Barbagallo}.
This is due to the enhanced,
conduction-electron-mediated RKKY coupling between the Eu 4\textit{f}
spins \cite{Mauger86,Arnold08}.
In thin films and interfaces these fundamental magnetic properties can also be influenced by additional factors,
such as surface-induced modification of the crystalline environment
and of the band structure \cite{Vaz} as well as magnetic proximity
effects \cite{Sperl,Maccherozzi,Menshov}. These interface effects have been studied experimentally mainly in 3$\textit{d}$
systems, be they itinerant ferromagnets \cite{Vaz} or transition metal oxides
\cite{Okamoto,Freeland}, while interfaces of the 4$\textit{f}$ compound EuO
have only been analyzed theoretically \cite{Lee,Wang}.

In this Letter we study systematically the Curie
temperature $T_C(d)$ and magnetic moment per Eu atom, $m(d)$, in
dependence of the thickness $d$ of ultrathin layers of oxygen-deficient
EuO$_{1\mbox{-}x}$ in the thickness range from 2 to 6 nm,
interfaced with Pt capping layers. Our previous investigation in the range from 7 to 12 nm for various O-defect concentrations
\cite{Barbagallo} did not show a thickness dependent variation of the magnetic properties
relative to bulk samples. However, in the ultrathin
layers of the present work we find a systematic reduction of both $T_C(d)$ and
$m(d)$ with decreasing $d$. Our analysis shows that band bending and the reduced number of nearest neighbors at the interface cause the thickness-dependent functionality of $T_C(d)$ and $m(d)$, due to the increased relative importance of the interface. We are then able to estimate the charge screening length and the spatial
extension of the RKKY interaction in EuO$_{1\mbox{-}x}$.

\begin{figure}[t]
\vspace{-30pt}
\includegraphics[width=0.47\textwidth]{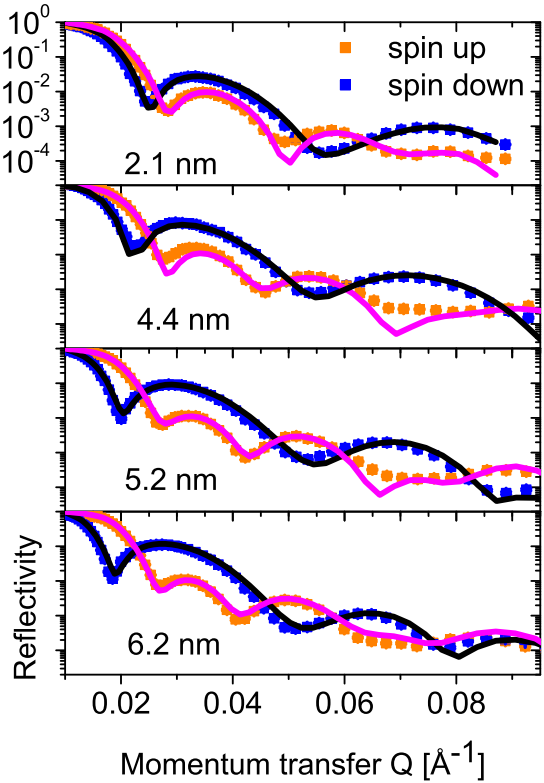}
\vspace{-20pt}
\caption{\label{fig:1}(Color online) PNR data (datapoints) and fit (lines) for
varying thickness, $\textit{d}_{\textsc{PNR}}$=2.1, 4.4, 5.2 and 6.2 nm.
Data taken at T=5 K with an applied magnetic field of 3 kOe.}
\end{figure}

Thin films of EuO$_{1\mbox{-}x}$ with x=4\% were deposited by co-sputtering
of Eu$_{2}$O$_3$ and Eu on Si
substrates with a Pt buffer and capping layer of 10 nm each, as described in
Ref. \cite{Barbagallo}. The samples were characterized by superconducting
quantum interference device (SQUID), x-ray reflectometry (XRR) and polarized
neutron reflectometry (PNR) on the CRISP beamline at ISIS \cite{isis},
following the same analysis as carried out in Ref.~\cite{Barbagallo}.
Since EuO is unstable towards phase separation, i.e. towards formation of elemental Eu and of the nonmagnetic, chemically
stable oxide Eu$_{2}$O$_{3}$, it is crucial to determine the number density of
magnetic Eu atoms in order to measure the magnetic moment per
Eu atom in the EuO phase, $m(d)$. This is achieved by fitting the PNR
data to a theoretical model with the following parameters: neutron scattering length, neutron absorption, atom number density, fraction of nonmagnetic phases,
magnetic film thickness $d$, and total magnetic moment of each layer, see Ref.~\cite{Barbagallo}.
The PNR data and theoretical fits are shown in Fig.~\ref{fig:1}, where the reduction in peak spacing and the progressive separation of the spin
up and spin down curves track the increase in thickness.
The other fit parameters were found to be the same
as for the thicker samples previously
measured \cite{Barbagallo}. The samples were polycrystalline and the
interlayer roughness was estimated to be about 0.6 nm (rms amplitude).

\begin{figure}[b]
\vspace{-10pt}
\includegraphics[width=0.47\textwidth]{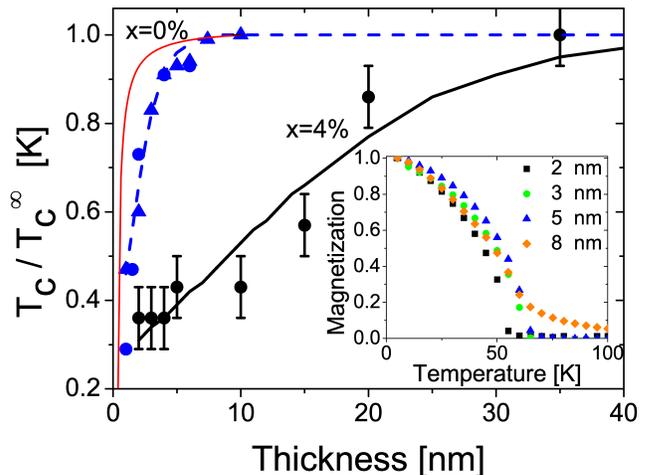}
\vspace{-10pt}
\caption{\label{fig:2}
(Color online) T$_{C}$ / T$_{C}$$^{\infty}$ of EuO$_{1\mbox{-}x}$
for varying thickness, for $\textit{x}$=4\% (black circles). The data
for $\textit{x}$=0\% (blue triangles and circles) is taken from Refs.
\cite{Santos,Muller}. Above 40 nm T$_{C}$ for $\textit{x}$=4\% saturates to 140
K (normalized value of unity, datapoints not shown). The red curve
represents Eq.~(\ref{eq:1}) normalized to its value for the largest
experimentally considered thickness (10 nm). The black and blue fit lines are
described in the text. The inset shows the magnetization as a function of
temperature for increasing thickness, with a 50 Oe applied field. Each magnetization curve is normalized to its own value at 5 K.}
\end{figure}
{\it Curie temperature.}
The thickness-dependent $T_{C}$ for thin films of EuO$_{0.96}$ in the thickness range between
2 and 40 nm is plotted in Fig.~\ref{fig:2}, normalized to the respective bulk value
$T_{C}^{\infty}$, together with data for EuO taken
from Refs. \cite{Santos,Muller}.
The reduction of T$_{C}$ for stoichiometric, i.e. insulating EuO can be
understood qualitatively by describing the Eu 4$\textit{f}$ subsystem within a spin $S=7/2$ Heisenberg model
with an effective nearest-neighbor spin exchange
coupling $J$ \cite{Arnold08}.
In mean field theory $T_C^{(MF)} = {ZJ}/{4k_B}$, that is
proportional to the number of nearest neighbors $Z$,
which is reduced from $Z_b=12$ in the bulk fcc lattice of EuO to
$Z_i=8$ at the interface ($J$ is the spin exchange coupling and $k_B$ the Boltzmann constant). Thus, averaging $Z$ over a film with $n$ atomic layers (a monolayer of EuO is 0.25 nm) and two interfaces leads to a reduction in $T_C$:
\begin{equation}
T_C^{(MF)} = \frac{2Z_i+(n-2)Z_b}{n}\ \frac{J}{4k_B}\ , \quad n\geq 2.\label{eq:1}
\end{equation}
This expression, after normalization to $T_C^{\infty}$ free of adjustable
parameters, is shown in
Fig.~\ref{fig:2} (red curve). We attribute the stronger experimental $T_C$ suppression to the fact that
in EuO next-nearest-neighbor couplings are not negligible \cite{Arnold08}
and that in thin films fluctuations, not included in mean field theory,
become increasingly important.

For oxygen-deficient EuO$_{0.96}$ the simple analysis in terms of a reduced number of neighbors is not sufficient due to the additional exchange interaction pathway mediated by the occupied conduction-band states, that is long-range RKKY. While the the RKKY interaction causes an increase in the absolute value of $T_C$ relative to undoped EuO \cite{Arnold08},
it also has the consequence of making the films more susceptible to surface effects: the reduction in $T_C$ extends up to significantly larger thicknesses compared to undoped EuO.
We can extract an experimental estimate for the range $\xi$ of the
effective spin coupling by fitting the experimental data with a phenomenological
Fermi-like function,
$T_C/T_C^{\infty}= [exp(1-d/\xi)+1]^{-1}$, which describes both the
$T_C$ saturation for large $d$ and the approximately linear $T_C$ suppression for
small $d$, by a single length scale $\xi$. The
best fits yield an effective range of $\xi\approx 1.2$ nm for
$x=0$\% and $\xi\approx 9$ nm for $x=4$\%.
We also note that we have attempted to reproduce the reduction in $T_C$
for EuO$_{0.96}$ by performing mean field calculations for a layered Heisenberg model
with an additional, RKKY-induced, effective spin exchange coupling
$J'(z)=J_0 \cos( 2 k_F z)/z$, cut off at the thermal length,
where $z$ is the distance between two spins
perpendicular to the interface. We found a
$T_C$ reduction for films up to $d=40$ nm (not shown) in qualitative agreement with
experiment; However, it was not possible to reproduce the experimental $T_C(d)$
curve quantitatively by a spatially constant strength $J_0$ of the RKKY-induced interaction.
We attribute this to our finding,
which will be discussed below in the context of the magnetic moment data,
that a substantial conduction band bending occurs near the interfaces,
leading to spatially dependent modifications of the RKKY interaction.

\begin{table}[t,floatfix]
\caption{\label{tab:1}Magnetic moment of EuO$_{0.96}$ with varying
thickness.}
\begin{ruledtabular}
\begin{tabular} {cccc}
$\textit{d}_{\textsc{PNR}}$ & $\textit{d}_{\textsc{XRR}}$ & $\mu(\textit{B}$)
PNR ($\pm{0.09}$ $\mu_{\textsc{B}}$)\\
$\text{(nm)}$  &$\text{(nm)}$ & ($\mu_{\textsc{B}}$/Eu atom)\\
\colrule
2.1&1.8&6.41\\
4.4&4.5&6.80\\
5.2&5.5&6.99\\
6.2&5.8&7.08\\
11.7&12.3&7.07\\
\end{tabular}
\label{Table1}
\end{ruledtabular}
\end{table}

\begin{figure}[b]

\vspace{-20pt}
\begin{centering}
\includegraphics[width=0.53\textwidth]{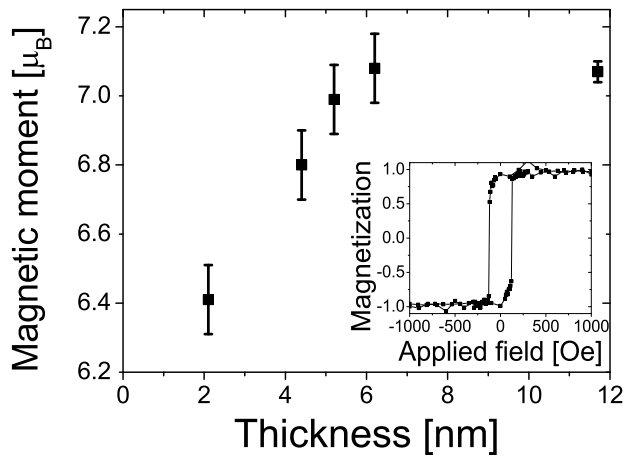}
\end{centering}
\vspace{-10pt}
\caption{\label{fig:3}Magnetic moment of EuO$_{0.96}$ for varying
thickness, $\textit{t}_{\textsc{PNR}}$=2.1, 4.4, 5.2 and 6.2 nm. The 11.7 nm
datapoint is taken from Ref. \cite{Barbagallo}.
The inset shows a hysteresis loop taken by SQUID at 5 K for
a 2 nm sample.}
\end{figure}

{\it Magnetic moment reduction.}
The magnetic moment per Eu atom (measured by PNR at 5 K, c.f. Fig.~\ref{fig:1})
and thickness (measured by PNR and XRR) for magnetically saturated
EuO$_{0.96}$ films are indicated in Table~\ref{tab:1} (the data for the 11.7 nm sample is taken from Ref. \cite{Barbagallo}).
The expected magnetic moment for EuO$_{0.96}$ is 7.08 $\mu_{\textsc{B}}$/Eu
atom \cite{Barbagallo}. As seen in Fig.~\ref{fig:3} there is
a marked, approximately linear reduction of the magnetic moment with
decreasing thickness, by up to 9\% for the 2.1 nm sample.
We note that a decreased value for the moment of ultra thin films of
stoichiometric EuO is visible in the data reported by Santos $\textit{et al.}$
\cite{Santos}, which display a monotonic decrease of the magnetic moment with decreasing
thickness \footnote[1]{The fact that their measured moment per Eu atom of 7.4 $\mu_{\textsc{B}}$
in the 6 nm film exceeds the maximum possible value for stoichiometric EuO is
attributed to an overall underestimation of the film thickness.}.

To understand the moment reduction, several effects must be considered.
In contrast to the $T_C$ suppression, the moment reduction
cannot be understood in terms of surface-reduced effective exchange couplings,
since the saturated ferromagnetic moment is independent of the coupling.
Further, we can exclude that the origin of the moment reduction lies in a
renormalization of the Land\'e factor $g$ near the interface,
to a value significantly
below the bulk one: In saturated thicker (bulk-like) films of EuO$_{1\mbox{-}x}$ with $\textit{x}$=4\% the observed
magnetic moment per Eu atom is $m$ = 7.08\ $\mu_{\textsc{B}}$
(c.f. Table~\ref{tab:1}), i.e. equal to the combined maximum spin moment
of the Eu 4$\textit{f}$ electrons and of the two dopant electrons per O-defect,
$m = g\ (7/2 + x)\ \mu_{\textsc{B}}$, where the Land\'e factor assumes its
vacuum value, $g=2$. This indicates that orbital, band structure or
many-body effects do not play a significant role for the $g$-factor.
Hence we do not expect that $g$ is modified due to orbital quenching
or a change of band structure near the interface. Moreover, the
Land\'e factor of Pt has consistently been reported
to be larger than 2 \cite{Gustafsson86}, so that we don't expect a reduction
of $g$ in EuO$_{1\mbox{-}x}$ due to proximity to the Pt capping layer. The
moment reduction in thin EuO$_{1\mbox{-}x}$ films cannot thus be explained by
a reduction of the $g$-factor.
Another factor to consider is that the 4$\textit{f}$ orbitals of EuO have been reported to be susceptible to pinning from the local crystalline
environment \cite{Laan,Arenholz}; this could cause a reduction of the moment. However, such surface pinning would result in
a strongly non-linear thickness dependence of the average moment, with a functionality of the type
given by Eq.~(\ref{eq:1}) (red curve in Fig.~\ref{fig:2}). Since this fails to
fit the measured behavior which is linear in the
thickness range of $d=2 - 6$~ nm, as seen in Fig.~\ref{fig:3}, we can exclude pinning effects as the primary cause of the observed
behavior.

The last factor we consider in our analysis, and the one with definite support from the experimental data, is the large difference between the work functions $\Phi$ of EuO$_{1\mbox{-}x}$
($\Phi_{EuO}\lesssim 1$~eV \cite{Eastman69}) and of the Pt capping layer ($\Phi_{Pt}\approx 5.6$~eV). One expects a significant electron transfer
from the EuO$_{1\mbox{-}x}$ film to the Pt layer, resulting in an interface
potential $V(z)$, where $z$ is the vertical distance from the interface.
Since the difference between the work functions, $\Phi_{Pt}-\Phi_{EuO} \approx 4.6$~eV,
is larger than the binding energy of the Eu $4f$ band, whose upper edge
lies about 1.2~eV below the conduction band \cite{Eastman69,Steeneken02}, the charge transfer will involve not only the bulk conduction electrons, but also the $4f$ electrons, i.e. the Eu $4f$ band will be bent upward to cross the Fermi level.
Therefore, we expect that an important origin of the moment reduction below
the Hund's rule moment of $m_{4f}= 7\ \mu_{\textsc{B}}$ for elemental Eu
is that near the interface the Eu $4f$ orbitals are partially depopulated
into the Pt capping layers by an upward bending of the conduction band
as well as of the Eu $4f$
valence band, such that the latter crosses the chemical potential.
The length scale for the range of the band bending is controlled by the charge
screening lengths in the EuO conduction and $4f$ bands, i.e.
significantly shorter than the range of the RKKY interaction. This is
consistent with our experimental finding
that the bulk moment is recovered for film thickness $d\geq 6$ nm
(Fig.~\ref{fig:3}), while the bulk $T_C$ is obtained only for
thickness $d\geq 40$ nm (Fig.~\ref{fig:2}).
Signatures of a modified surface electronic structure have been reported
previously for EuO bulk crystals \cite{Sattler,Meier} and possibly thin
films \cite{Song}. Direct experimental evidence for a significant
upward band bending at the Pt-EuO$_{1-x}$ interface can, however, be seen
in the temperature dependence of the magnetization shown in the inset of
Fig.~\ref{fig:2}. These magnetization curves are notable for the
gradual disappearance of the secondary dome with decreasing film thickness.
This dome is always present in the magnetization curves of doped bulk
EuO samples \cite{Schmehl,Barbagallo}; it constitutes a deviation from the Brillouin function of a Heisenberg lattice of localized spins \cite{Barbagallo} and it is
associated with the magnetization of the occupied conduction band
\cite{Arnold08}. The disappearance of the secondary dome thus indicates
a complete depopulation of the EuO$_{1\mbox{-}x}$ conduction band caused by an upward bending above the chemical potential.
Thus, we conclude that surface band bending is the most significant
cause for the moment reduction in thin EuO$_{1\mbox{-}x}$ films.
The band bending effect due to interface charge transfer can be
calculated within a semiclassical Thomas-Fermi theory, where the local
interface potential, $V(z)$, and the interface charge density distribution
in that potential, $\delta\rho(z)$, are calculated self-consistently
using Gauss' law and the locally shifted band energy levels, respectively,
and minimizing the total energy of the interface
\footnote[2]{M. Barbagallo, T. Stollenwerk, and J. Kroha, in preparation.}.

To conclude, we have performed systematic measurements of
the Curie temperature and layer-average magnetic moment in thin,
oxygen-deficient EuO films in dependence of the film thickness.
These measurements enabled us to study the influence of the film interface
on these quantities and to analyze the physical effects contributing to
their reduction. In stoichiometric EuO the Curie temperature is reduced for film thicknesses smaller than 10 nm,
and we found that this reduction
can be well explained semiquantitatively by the reduced number of
neighboring magnetic atoms at the surface of the Eu sublattice.
In electron doped, i.e., semiconducting EuO$_{0.96}$ there is an overall, numerical enhancement of the Curie temperature
with respect to stoichiometric EuO,
but the surface-induced reduction extends up to higher film thicknesses
of about 40 nm. The overall absolute-value enhancement and the
thickness-dependent reduction can both be understood qualitatively in terms of the
conduction-electron mediated, long-range RKKY spin-exchange interaction
operative in these conducting films.
Analyzing the reduction of the layer-average magnetic moment per Eu atom, we found
interface-induced band bending and concomitant partial depopulation of the
Eu $4f$ band to be the dominant mechanism, besides possible pinning effects.
This conclusion was reached by estimating the energy gain for transferring
Eu $4f$ electrons at the interface with the Pt capping layers.
Our results are also relevant for applications of EuO
in spintronics, where interface effects naturally play
an important role. Especially, our conjecture that the
moment may be controlled by the interface work functions
may be useful for maximizing the magnetic moment of EuO in ultrathin layer structures.
More detailed, spatially dependent calculations will, however, be needed
to understand the reduction of the magnetic moment as well as of the
Curie temperature quantitatively.

\begin{acknowledgments}We would like to thank the STFC for funding the PNR
measurements.
This work was financially supported in part (J.K, T.S.) by the Deutsche
Forschungsgemeinschaft through SFB 608.
\end{acknowledgments}

\bibliography{EuO_thinfilm}

\end{document}